\documentclass[reprint,aps]{revtex4-2}

\usepackage{graphicx}% Include figure files
\usepackage{dcolumn}% Align table columns on decimal point

\usepackage[utf8]{inputenc}
\usepackage{siunitx}
\usepackage[german, english]{babel}

\begin{document}

%\preprint{}

\title[Demystifying dust contamination in quantum optics labs]{Demystifying dust contamination in quantum optics labs:\\ measurements and recommendations}

\author{Jonas Gottschalk}
  \affiliation{Physikalisches Institut, Universität Bonn, Nussallee 10, 53115 Bonn, Germany}
\author{Simon Stellmer}
  \email{stellmer@uni-bonn.de}
  \affiliation{Physikalisches Institut, Universität Bonn, Nussallee 10, 53115 Bonn, Germany}

\date{\today}

\begin{abstract}

Experiments in the field of quantum optics often require very low concentrations of dust particles in the laboratory, but the complexity of working routines precludes operation within a proper clean room. Research teams have established a multitude of different approaches, precaution measures, and habits to keep the delicate optics setups free of contamination. Here, we systematically quantify dust particle concentration during day-to-day operation of a quantum optics lab, assess the effectiveness of various measures, and give practical recommendations.

\end{abstract}

\pacs{}% insert suggested PACS numbers in braces on next line

\maketitle

\section{Introduction}

Quantum optics laboratories are often centered around delicate optics mounted on an optical table. The optics need to be protected from dust, which would otherwise lead to absorption and thus reduce the transmission and reflection of optical surfaces. Prominent examples include the degradation of the finesse of an optical resonator, simple burning of optical coatings due to heating from high-intensity lasers, and attraction of dust onto optical surfaces through the optical tweezer effect.

Aside from experiments involving high-power lasers, such setups are rarely operated in proper clean rooms. The complexity of the setups, the number of persons working on them often, and the lack of training often interfere with established clean room environments. Instead, a miniaturized clean room enclosure is set up above the optical table, and certain procedures are established to reach a practical compromise between air quality and ease of operation.

A multitude of habits have developed in hundreds of optics labs worldwide. Some require the usage of dedicated lab shoes or coats, some don't. Some feature ``Strictly no soldering in this room" or ``Don't tear paper" signs, while other are covered with sticky mats. In this work, we seek to remove some of the voodoo around the topic of dust contamination, and we will give practical recommendations to keep dust contamination low.

\section{\label{sec1}Study design, setup, and sensors}

For our studies, we subdivide the laboratory into two regions: the room itself and the enclosed optical table underneath an air filter unit.

\subsection{Laboratory}

The laboratory used for these studies has a size of 20 sqm and hosts two optical tables. An air conditioning system keeps the temperature near \SI{21}{\celsius}. The humidity is not controlled, and there is no fresh air supply except through the entrance door. To measure the dust concentration in the laboratory the Sensirion SEN55 Environmental Sensor Node, referred to below simply as Sen55, is used.
The Sen55 was chosen because of its low cost, the ability to distinguish between different particle sizes and its compatibility with an Arduino micro-controller board, used for communication with the sensor. Besides the dust concentration, the temperature, pressure, and humidity are monitored as well.

Optical particle counters, as the Sen55, work on the principle of laser scattering. Air is sucked in by a fan and guided through a detection volume where the particles scatter the light of a laser that is oriented perpendicular to the direction of air flow. The main process here is that of Mie scattering. A photo diode then detects the scattered light \cite{GailLothar2018R}.
Particle sizes and concentrations are calculated by electronic processing of the photo diode signal and the known value of air flow \cite{Pedersini}. The Sen55 uses a \SI{660}{\nano\meter} laser, it can distinguish particle sizes and bins the count rates into six different size ranges. In all our studies, we find the size distribution of particles to be constant. For simplicity, we will limit our analysis to particle concentrations for the size range between range between \SI{0.3}{\micro\meter} and \SI{2.5}{\micro\meter}. The Sen55 device outputs the particle concentration in \si{particles/\centi\metre^3} and corresponding PM values in \si{\micro\gram/\metre^3}  once every second.

The Sen55 sensors are calibrated to reference sensors by the manufacturer and then compared to each other in batches. The calibration process is described in Ref.\cite{sencalib}. To benchmark the performance of the Sen55 sensors, we place two of them in a tightly sealed plastic container. Particle concentrations obtained by the two sensors agree to within \SI{2}{\percent} and give constant readings after an equilibration time of about one day.

\subsection{Enclosure}

Our micro clean room enclosure is centered around an optical table of $1.2$ by \SI{2.4}{\metre} size, which contains the delicate optics. \SI{97}{\centi\metre} above the table, we install a panel made of aluminum frames and plastic boards. This panel hosts a 1210 by 600 mm ULPA airfilter (``flowbox") with an adjustable throughput of up to \SI{1400}{\metre^3/\hour}. The filter removes 99.9995\% of particles with a size of 0.12 $\mu$m, rated as U15 and qualified for clean room class 10/100. At a throughput of \SI{1100}{\metre^3/\hour}, the air flow is 0.45 m/s. The panel extends about 5 cm beyond the optical table on all sides, and is equipped with a total of 6 overlapping black foils (``curtains") that form the sides of the enclosure: one on each short side and two on each long side. The air supplied by the filter can leave the enclosure through the gap between the curtains and the optical table.

To measure dust concentrations within the enclosure, we use the Lasair III Aerosol Particle Counter. This sensor is calibrated and cleanroom certified and should only be operated in a comparably clean environment. The device can measure particles between \SI{0.3}{\micro\meter} and \SI{10}{\micro\meter} and categorizes them into six size intervals. Again, we will consider only the interval of smallest sizes. The Lasair sensor has a specified background of up to \SI{7}{particles/\meter^3}: this is the minimum concentration to which it can credibly measure dust contamination, corresponding to clean room class ISO 2 or better. 

As compared to the Sen55 device, the Lasair sensor has a higher air throughput and thus a higher sensitivity, it shows a smaller background and allows the user to adjust the averaging time. We perform side-by-side measurements of the Sen55 and Lasair sensors in a regime of intermediate dust concentrations. The readings differ by a factor of up to three.

\section{\label{sec2}Dust within the laboratory}

\begin{figure}
    \includegraphics[width=0.45\textwidth]{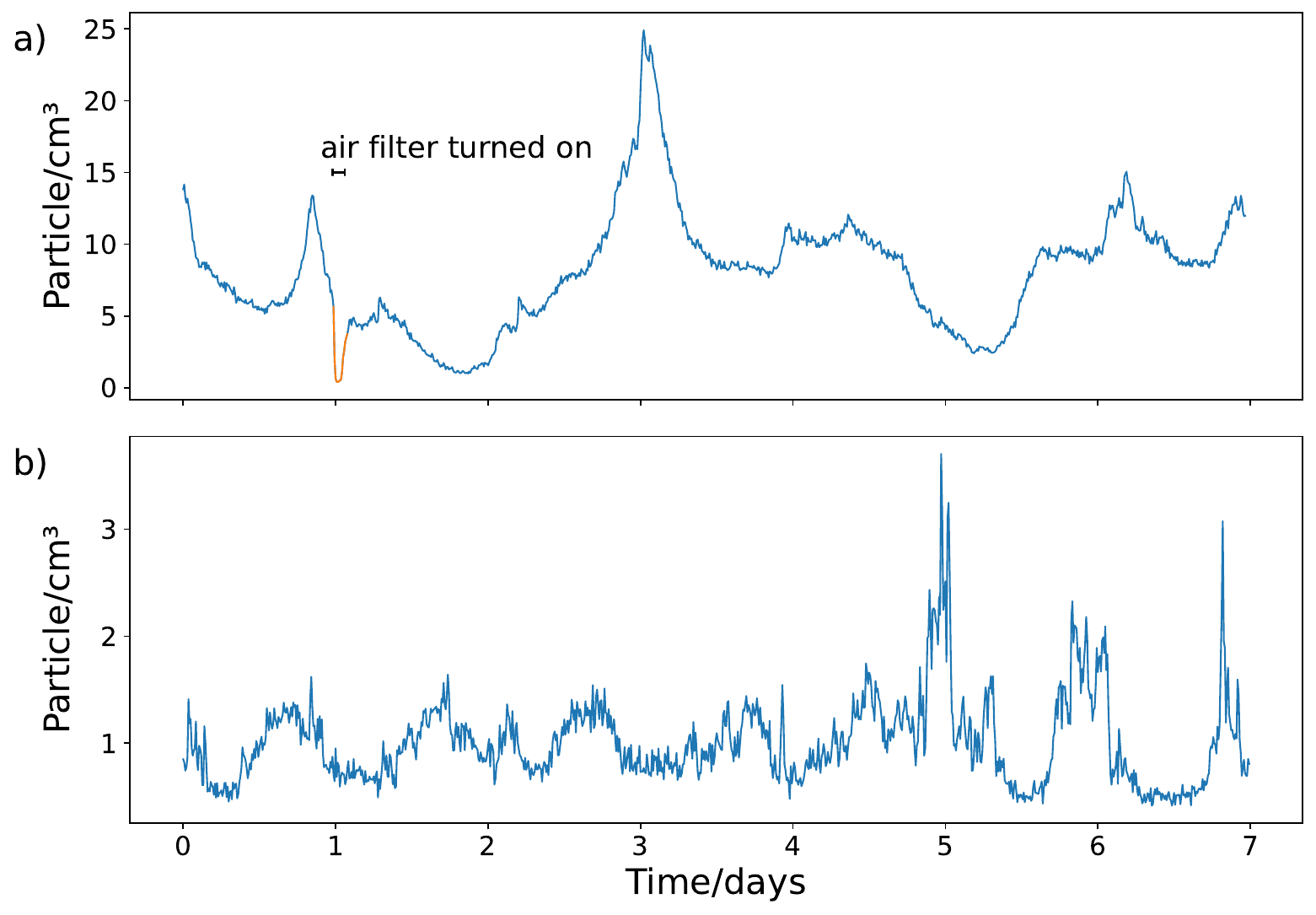}
    \caption{\label{fig:long_both} Particle concentrations between 0.3 and \SI{2.5}{\micro\meter} measured over two weeks without (a) and with (b) air filter units running.}
\end{figure}

We first measure the particle concentration in the laboratory without running the air filter unit. The data, taken with the Sen55 sensor over the course of one week, in shown in Fig.~\ref{fig:long_both}(a). During this time, we varied the temperature by $\Delta T = \SI{2.6}{\celsius}$ and the humidity by $\Delta H = \SI{7.8}{\percent}$, but find no correlation with particle concentration. Similarly, pressure variations of $\Delta P = \SI{25}{\hecto\pascal}$ show no correlations with particle concentration. The average value is \SI{8.1}{particles/\centi\meter^3}, where numbers range from \SI{1}{particles/\centi\meter^3} to \SI{25}{particles/\centi\meter^3} with no apparent pattern. In particular, there is no daily periodicity. 

The number concentrations correspond to a mean PM2.5 value (total mass of particles with size above 2.5\,$\mu$m per volume) of \SI{1.7}{\micro\gram/\centi\meter^3}. This value is a factor of about 10 smaller than the typical dust contamination in office buildings, where mean PM2.5 values between \SI{9.2}{\micro\gram/\centi\meter^3} and \SI{16}{\micro\gram/\centi\meter^3} have been reported \cite{MANDIN2017169}.

At the beginning of day 1, we turned the air filter unit on for a short time. An immediate reduction in particle concentration by about an order of magnitude can be observed, indicated by an arrow in Fig.~\ref{fig:long_both}(a). The $1/e$ time constant of reduction is about \SI[separate-uncertainty = true]{8(2)}{\minute} and agrees approximately with the time required to filter the air volume of the laboratory once (about 3\,min). 

Figure \ref*{fig:long_both}(b) shows a one-week measurement with the air filter unit constantly running. Here, the mean value was \SI{1.0}{particles/\centi\meter^3}, with values ranging between \SI{0.4}{particles/\centi\meter^3} and \SI{3.7}{particles/\centi\meter^3}. On average, the air filters reduce particle concentrations by \SI{88}{\percent}, such that the dust concentration in the laboratory is about two orders of magnitude lower than in a conventional office. Again, no correlation with ambient conditions was found. 
The peaks at the beginning of days 5 and 7 can not be assigned to specific events. In particular and maybe surprisingly, our data does not show signatures of persons working in the room.

We also measure particle concentrations at different heights in the room, but find no significant dependence.

\textit{Vacuum cleaning} -- We repeatedly run a vacuum cleaner robot through the room. No effect, neither positive or negative, can be observed. We conclude that the dust accumulating on the floor (macroscopic particles such as sand) is entirely decoupled from the class of micrometer-sized particles relevant for our studies.

\begin{figure}
    \includegraphics[width=0.45\textwidth]{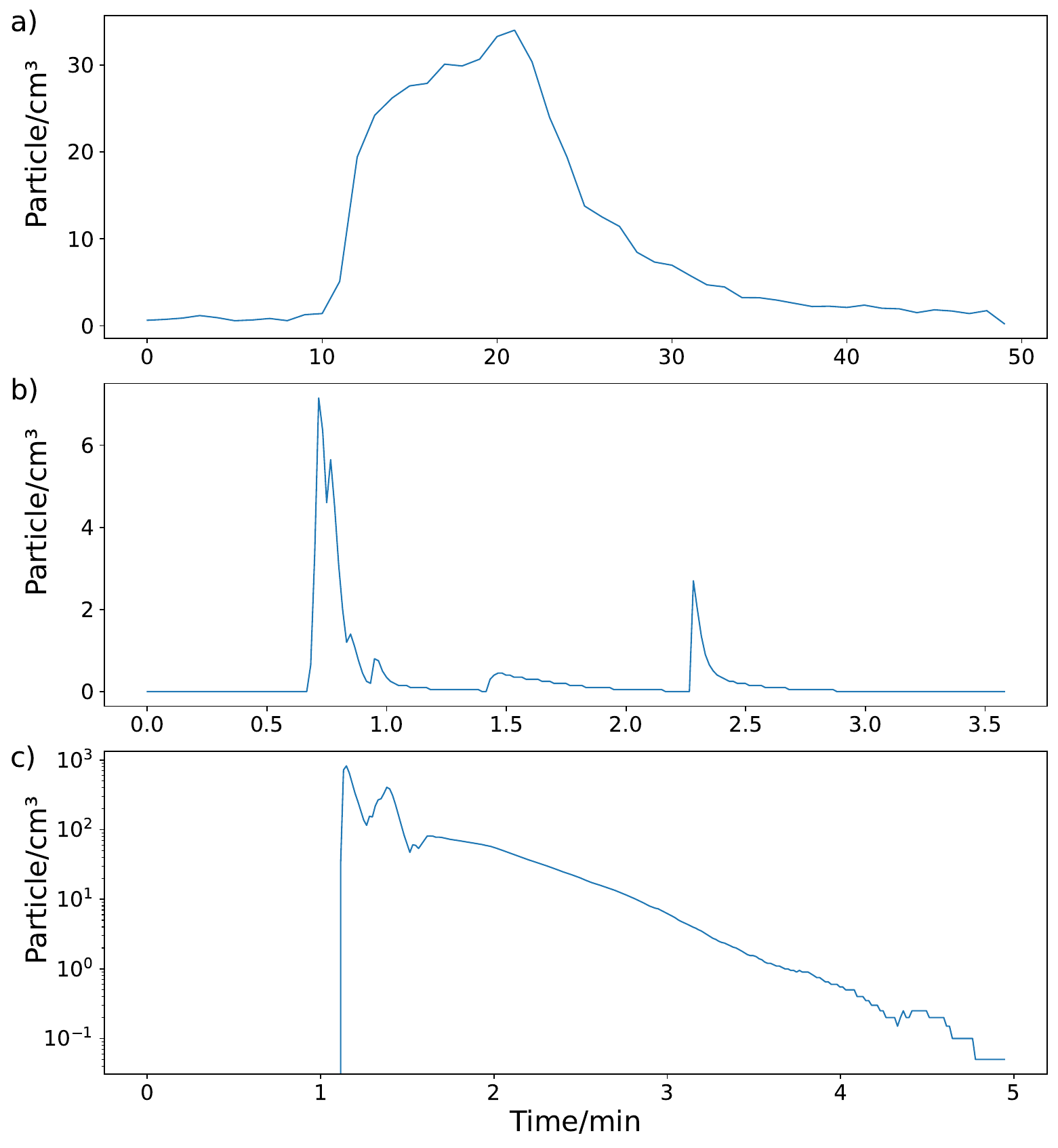}
    \caption{\label{fig:wasmachtesschlechter} (a) Venting the room for \SI{10}{\minute} leads to a drastic increase of particle concentration. (b) Tearing a piece of cardboard leads to an increase in particle concentration. (c) Melting of a \SI{5}{\centi\meter} piece of soldering tin increases particle concentration dramatically.}
\end{figure}

\textit{Venting the room} -- We vent the room for \SI{10}{\minute} by opening a window. The particle concentration increases from below \SI{1}{particles/\centi\meter^3} to above \SI{30}{particles/\centi\meter^3}, and returns to the original value within half an hour; see Fig.~\ref*{fig:wasmachtesschlechter}(a). We conclude that unfiltered air from outside contains significant amounts of microscopic particles (dust, pollen) that deteriorate air quality considerably.

\textit{Tearing of cardboard} -- Similarly, it is known that the tearing of paper and cardboard frees large amounts of microscopic pieces of fiber. Here, we tear pieces of cardboard and observe an increase in particle concentration by a factor of about ten; see Fig~\ref*{fig:wasmachtesschlechter}(b). Using a knife to cut the cardboard does not increase the measured particle concentration at all.

\textit{Soldering} -- Soldering is known to generate smoke that might settle on the optics. Within an enclosed environment, we use a conventional soldering iron to melt about \SI{5}{\centi\meter} of soldering tin about \SI{20}{\centi\metre} away from the Sen55 sensor. The data, presented in Fig.~\ref*{fig:wasmachtesschlechter}(c), shows the massive emission of microscopic particles. 

\textit{Persons in the room} -- We repeatedly probed for the influence of persons in the room on the overall dust contamination in the laboratory, but could not observe a significant impact.

\section{\label{sec3}Dust within the flow box enclosure}

\begin{figure*}[ht]
    \includegraphics[width=1.0\textwidth]{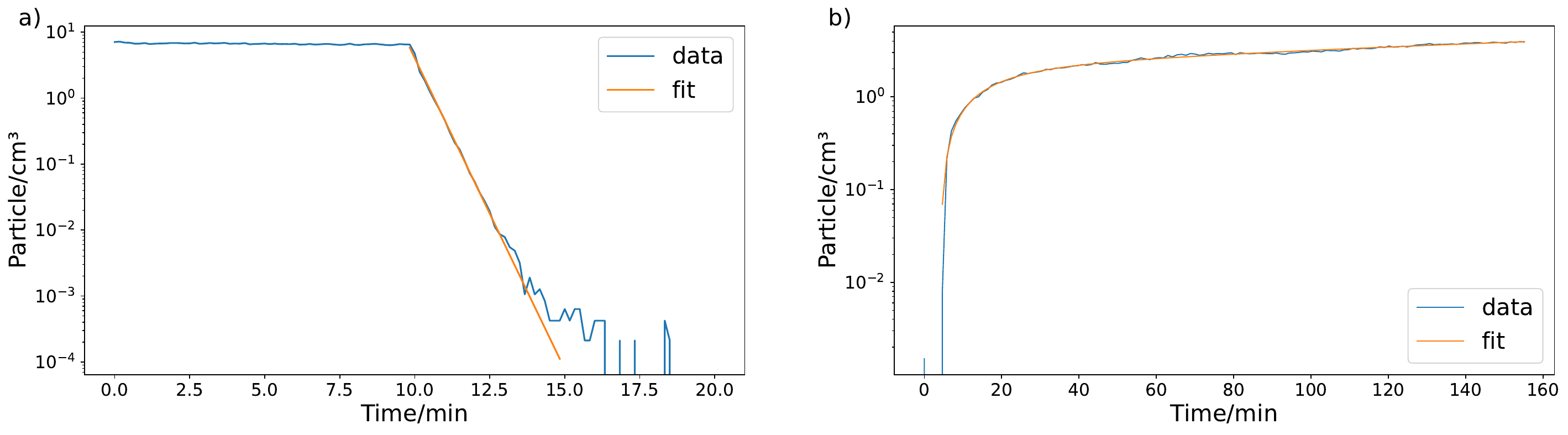}
    \caption{\label{fig:filteranaus_nichtlog} Turning the air filter unit on and off again.}
\end{figure*}
We now turn to measurements within the enclosure below the flow box, which were performed with the Lasair particle counter.
The Lasair is set here to measure the dust concentration over \SI{10}{\second} for each point.
The effect of the air filters is shown in Fig.~\ref*{fig:filteranaus_nichtlog}(a): at a time constant of \SI[separate-uncertainty = true]{27.6(0.2)}{\second}, the particle concentration decreases exponentially by at least four orders of magnitude and falls well below the detection threshold of the chosen integration time. To assess the steady-state particle concentration, we increase the measurement interval to 2 days and find a value of \SI{8.5}{particle/\metre^3}. This value is very close to the specified ``dark count" background of the sensor (\SI{7}{particle/\metre^3}) and indicates that the particle concentration indeed reaches ISO 1 (\SI{1}{particle/\metre^3} at a size $>\,3\mu$m) to ISO 2 (\SI{10}{particle/\metre^3}) standards.
% to an equilibrium value of about xx.

Turning the filter unit off again leads to a sudden increase in particle concentration that reaches the background value of the room after a few hours; see Fig.~\ref*{fig:filteranaus_nichtlog} (b). Here, the measured time constant is \SI[separate-uncertainty = true]{800(50)}{\second}.

\begin{figure}[h]
    \includegraphics[width=0.45\textwidth]{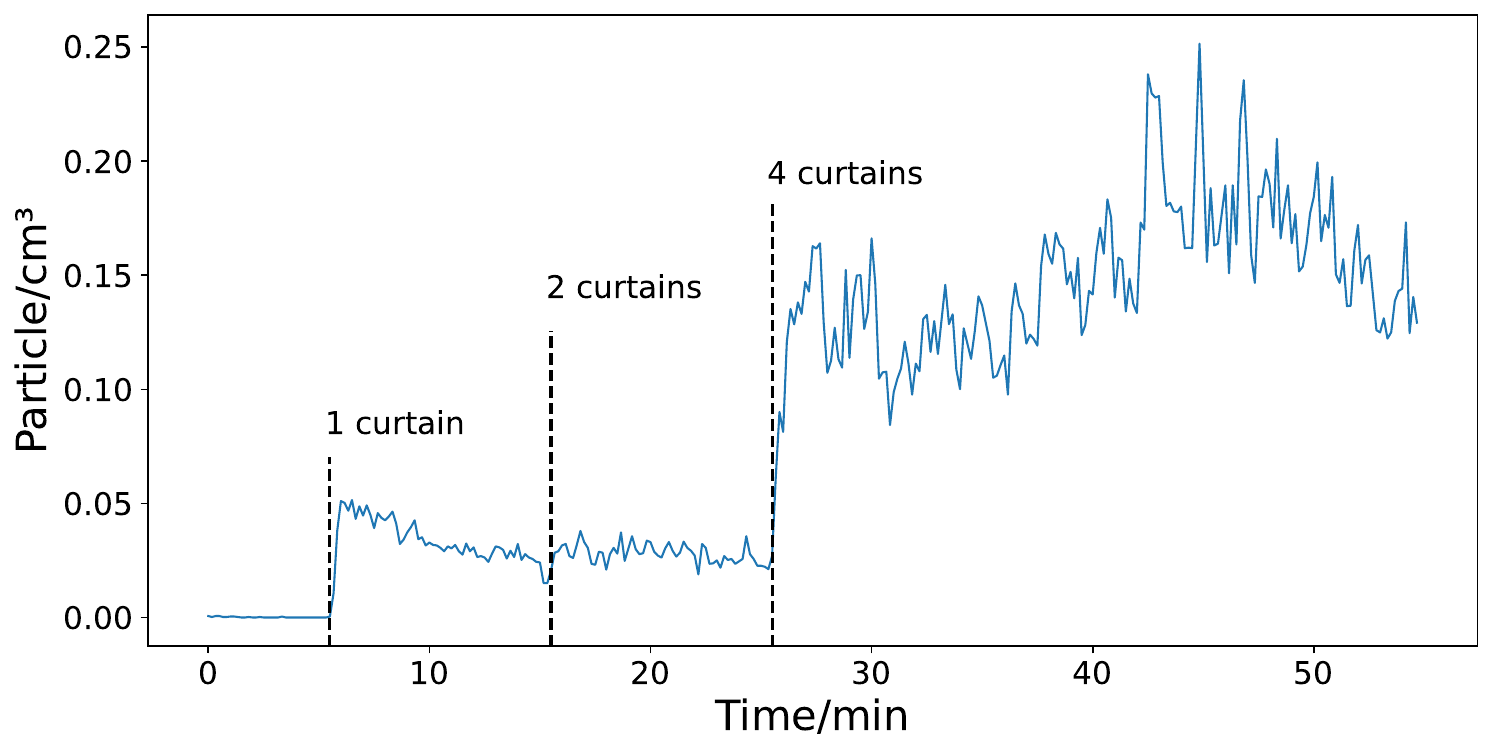}
    \caption{\label{fig:folienöffnen} Opening one curtain at a time (arrows) increases particle concentration.}
\end{figure}

\textit{Opening the curtains} -- Opening the curtains undermines the intended flow of air to keep particles from entering the enclosure. Here, we open the curtains one-by-one and observe the reduction in air quality. Opening one curtain on a long side of the table (about \SI{1.2}{\metre^2}, one-sixth of the total wall surface) increases the particle concentration from near-zero to about \SI{0.025}{particles/\centi\meter^3}, following an initial overshoot to about \SI{0.05}{particles/\centi\meter^3} possibly induced by swirl-up of dust. After \SI{10}{\minute}, we open the adjacent curtain on the long side of the optical table: no further increase is observed. After another \SI{10}{\minute}, we open the two curtains on the opposite side: the particle concentration increases considerably and settles between 0.1 and 0.25 \SI{}{particles/\centi\meter^3}; see Fig.~\ref{fig:folienöffnen}. We conclude that the opening of opposite sides allows substantial air flow, driven by air conditioning system, to pass over the optical table.

\textit{Using protective wear} -- To assess the influence of a person working within the enclosure at the optical table, we perform two measurements. In the first round, a person works under the completely closed flow box ``behind" the foil. The person is wearing lab shoes, but otherwise regular clothes. Within \SI{30}{\minute}, the count rate increases steadily and saturates around \SI{0.01}{particles/\centi\meter^3}. In the second round, the person is wearing a protective hood and a lab coat certified for clean rooms usage. While all other circumstances are kept unchanged, the particle concentration increases only to about \SI{0.001}{particles/\centi\meter^3}. We conclude that even simple protection measures can reduce human dust emission by almost an order of magnitude.

\section{\label{sec4}Conclusion and recommendations}

In conclusion, we find that the type of micro clean room studied here works very effectively and meet ISO 1 to ISO 2 standards. The filter unit must not be turned off, not even for a few minutes, as dust immediately enters the volume above the optical table. Once turned on again, the air filter can establish clean room conditions within a few tens of seconds.

Opening parts of the curtains can be tolerated surprisingly well, but substantial air flow across the table, enabled by opening curtains on opposite sides, should be avoided.

Venting the room (smoke or possibly pollen from outside), soldering, and tearing of paper all increase particle concentration considerably and should be avoided. Wearing a lab coat and a protective hood allows a person working at the optical table to reduce dust contamination by about an order of magnitude. Other measures such as sticky mats and vacuum cleaning or the deliberate increase of the humidity have no effect on the overall particle density in the laboratory. We find that the dust emission of persons is surprisingly small and can only be detected at small distances and long exposure times. Their presence has negligible influence on the dust contamination in the laboratory.

For the constant monitoring of particle concentrations, comparably cost-effective and easy-to-integrate sensors are well suited and reach the same performance as premium-class calibrated and clean room certified sensors.

\begin{acknowledgments}
We thank numerous research groups for discussions and for sharing their experience. In particular, we thank Selim Jochim (Heidelberg) for drawing our attention to this topic. This work received funding from Deutsche Forschungsgemeinschaft (DFG) through the Cluster of Excellence ML4Q (EXC 2004/1 – 390534769) and from the European Research Council (ERC) through project number 757386.
\end{acknowledgments}

\section*{Data Availability Statement}
The data that support the findings of this study are available from the corresponding author upon reasonable request.

\bibliography{Dust_contamination}

\begin{thebibliography}{4}%
\makeatletter
\providecommand \@ifxundefined [1]{%
 \@ifx{#1\undefined}
}%
\providecommand \@ifnum [1]{%
 \ifnum #1\expandafter \@firstoftwo
 \else \expandafter \@secondoftwo
 \fi
}%
\providecommand \@ifx [1]{%
 \ifx #1\expandafter \@firstoftwo
 \else \expandafter \@secondoftwo
 \fi
}%
\providecommand \natexlab [1]{#1}%
\providecommand \enquote  [1]{``#1''}%
\providecommand \bibnamefont  [1]{#1}%
\providecommand \bibfnamefont [1]{#1}%
\providecommand \citenamefont [1]{#1}%
\providecommand \href@noop [0]{\@secondoftwo}%
\providecommand \href [0]{\begingroup \@sanitize@url \@href}%
\providecommand \@href[1]{\@@startlink{#1}\@@href}%
\providecommand \@@href[1]{\endgroup#1\@@endlink}%
\providecommand \@sanitize@url [0]{\catcode `\\12\catcode `\$12\catcode `\&12\catcode `\#12\catcode `\^12\catcode `\_12\catcode `\%12\relax}%
\providecommand \@@startlink[1]{}%
\providecommand \@@endlink[0]{}%
\providecommand \url  [0]{\begingroup\@sanitize@url \@url }%
\providecommand \@url [1]{\endgroup\@href {#1}{\urlprefix }}%
\providecommand \urlprefix  [0]{URL }%
\providecommand \Eprint [0]{\href }%
\providecommand \doibase [0]{https://doi.org/}%
\providecommand \selectlanguage [0]{\@gobble}%
\providecommand \bibinfo  [0]{\@secondoftwo}%
\providecommand \bibfield  [0]{\@secondoftwo}%
\providecommand \translation [1]{[#1]}%
\providecommand \BibitemOpen [0]{}%
\providecommand \bibitemStop [0]{}%
\providecommand \bibitemNoStop [0]{.\EOS\space}%
\providecommand \EOS [0]{\spacefactor3000\relax}%
\providecommand \BibitemShut  [1]{\csname bibitem#1\endcsname}%
\let\auto@bib@innerbib\@empty
%</preamble>
\bibitem [{\citenamefont {Gail}\ and\ \citenamefont {Gommel}(2018)}]{GailLothar2018R}%
  \BibitemOpen
  \bibfield  {author} {\bibinfo {author} {\bibfnamefont {L.}~\bibnamefont {Gail}}\ and\ \bibinfo {author} {\bibfnamefont {U.}~\bibnamefont {Gommel}},\ }\href@noop {} {\emph {\bibinfo {title} {Reinraumtechnik}}},\ \bibinfo {edition} {4th}\ ed.,\ VDI-Buch\ (\bibinfo  {publisher} {Springer Nature},\ \bibinfo {address} {Berlin, Heidelberg},\ \bibinfo {year} {2018})\BibitemShut {NoStop}%
\bibitem [{\citenamefont {Pedersini}(2019)}]{Pedersini}%
  \BibitemOpen
  \bibfield  {author} {\bibinfo {author} {\bibfnamefont {F.}~\bibnamefont {Pedersini}},\ }\bibfield  {title} {\bibinfo {title} {Improving a commodity dust sensor to enable particle size analysis},\ }\href {https://doi.org/10.1109/TIM.2018.2834178} {\bibfield  {journal} {\bibinfo  {journal} {IEEE Transactions on Instrumentation and Measurement}\ }\textbf {\bibinfo {volume} {68}},\ \bibinfo {pages} {177} (\bibinfo {year} {2019})}\BibitemShut {NoStop}%
\bibitem [{sen(2020)}]{sencalib}%
  \BibitemOpen
  \href@noop {} {\emph {\bibinfo {title} {Sensor Specification Statement}}},\ \bibinfo {organization} {Sensirion AG} (\bibinfo {year} {2020}),\ \bibinfo {note} {version 1}\BibitemShut {NoStop}%
\bibitem [{\citenamefont {Mandin}\ \emph {et~al.}(2017)\citenamefont {Mandin}, \citenamefont {Trantallidi}, \citenamefont {Cattaneo}, \citenamefont {Canha}, \citenamefont {Mihucz}, \citenamefont {Szigeti}, \citenamefont {Mabilia}, \citenamefont {Perreca}, \citenamefont {Spinazzè}, \citenamefont {Fossati}, \citenamefont {{De Kluizenaar}}, \citenamefont {Cornelissen}, \citenamefont {Sakellaris}, \citenamefont {Saraga}, \citenamefont {Hänninen}, \citenamefont {{De Oliveira Fernandes}}, \citenamefont {Ventura}, \citenamefont {Wolkoff}, \citenamefont {Carrer},\ and\ \citenamefont {Bartzis}}]{MANDIN2017169}%
  \BibitemOpen
  \bibfield  {author} {\bibinfo {author} {\bibfnamefont {C.}~\bibnamefont {Mandin}}, \bibinfo {author} {\bibfnamefont {M.}~\bibnamefont {Trantallidi}}, \bibinfo {author} {\bibfnamefont {A.}~\bibnamefont {Cattaneo}}, \bibinfo {author} {\bibfnamefont {N.}~\bibnamefont {Canha}}, \bibinfo {author} {\bibfnamefont {V.~G.}\ \bibnamefont {Mihucz}}, \bibinfo {author} {\bibfnamefont {T.}~\bibnamefont {Szigeti}}, \bibinfo {author} {\bibfnamefont {R.}~\bibnamefont {Mabilia}}, \bibinfo {author} {\bibfnamefont {E.}~\bibnamefont {Perreca}}, \bibinfo {author} {\bibfnamefont {A.}~\bibnamefont {Spinazzè}}, \bibinfo {author} {\bibfnamefont {S.}~\bibnamefont {Fossati}}, \bibinfo {author} {\bibfnamefont {Y.}~\bibnamefont {{De Kluizenaar}}}, \bibinfo {author} {\bibfnamefont {E.}~\bibnamefont {Cornelissen}}, \bibinfo {author} {\bibfnamefont {I.}~\bibnamefont {Sakellaris}}, \bibinfo {author} {\bibfnamefont {D.}~\bibnamefont {Saraga}}, \bibinfo {author} {\bibfnamefont {O.}~\bibnamefont {Hänninen}}, \bibinfo {author} {\bibfnamefont
  {E.}~\bibnamefont {{De Oliveira Fernandes}}}, \bibinfo {author} {\bibfnamefont {G.}~\bibnamefont {Ventura}}, \bibinfo {author} {\bibfnamefont {P.}~\bibnamefont {Wolkoff}}, \bibinfo {author} {\bibfnamefont {P.}~\bibnamefont {Carrer}},\ and\ \bibinfo {author} {\bibfnamefont {J.}~\bibnamefont {Bartzis}},\ }\bibfield  {title} {\bibinfo {title} {Assessment of indoor air quality in office buildings across europe – the officair study},\ }\href@noop {} {\bibfield  {journal} {\bibinfo  {journal} {Science of The Total Environment}\ }\textbf {\bibinfo {volume} {579}},\ \bibinfo {pages} {169} (\bibinfo {year} {2017})}\BibitemShut {NoStop}%
\end{thebibliography}%

\end{document}